\documentclass[10pt]{article}

\usepackage[preprint]{tmlr}
\usepackage{xcolor}
\usepackage{amsmath}
\usepackage{amssymb}
\usepackage{array}
\usepackage{booktabs}
\usepackage{float}
\usepackage{microtype}
\usepackage{tabularx}
\usepackage{xurl}
\usepackage[
  colorlinks=true,
  linkcolor=blue,
  citecolor=blue,
  urlcolor=blue
]{hyperref}

\newcolumntype{Y}{>{\raggedright\arraybackslash}X}

\title{Open-Qwen-Music: An Auditable Framework for LLM-Based Music Composition and Diffusion Rendering}
\author{
  Yangbin Yu\thanks{Work done at Tencent.}
  \and
  Mingyu Yang\footnotemark[1]
}
\date{}

\begin{document}

\maketitle

\begin{abstract}
We present Open-Qwen-Music,
an open reconstruction of Qwen-Music and a
fully specified research
system for text-to-music generation that couples LLM-based semantic
composition with diffusion-based acoustic rendering. The system comprises a
25~Hz single-codebook music tokenizer, a 3B-parameter autoregressive Music LLM,
and a diffusion renderer producing 48~kHz stereo audio, following the
cross-module interfaces reported by Qwen-Music. The strongest systems of this
design remain closed, and prominent open music-generation projects
release weights and inference code without their training corpora or
end-to-end training implementations. This limits independent and controlled study of how information loss and
prediction errors propagate from semantic representation through
autoregressive planning to acoustic rendering.
To our knowledge, Open-Qwen-Music is the first fully open release of an
LLM-composition-plus-diffusion-rendering text-to-music system. Beyond model
weights and inference code, the release includes the training datasets and
provenance manifests, complete data-processing, annotation, training,
inference, and evaluation pipelines, configurations, and pretrained weights
for every learned module. Artifact manifests bind the identities of
these artifacts across the complete workflow.
Together, these artifacts establish a reproducible implementation of the modular architecture and provide an empirical basis for component-level analysis and future evaluation. We present the system as a transparent, executable research
baseline and a starting point for the community, not as evidence of quality
parity with Qwen-Music. Open-Qwen-Music is an ongoing effort, and we
will continue to improve its generation quality,
controllability, and robustness. All release artifacts are available at
\url{https://github.com/biang15343100-source/Open-Qwen-Music}.
\end{abstract}

\section{Introduction}
\label{sec:introduction}

Recent music-generation systems increasingly separate semantic composition
from acoustic rendering. A language model first plans long-range musical
content in a compact representation. A diffusion or flow model then renders
that representation into audio. Qwen-Music, ACE-Step~1.5, and LeVo~2 all
follow this broad design~\citep{xu2026qwenmusic,gong2026acestep,lei2026levo2}.
This decomposition offers a natural way to divide musical planning
from waveform synthesis, but it also creates coupled failure modes: information
discarded by the representation constrains the planner, and planner errors
become out-of-distribution conditions for the renderer. Studying these
interactions requires more than inference access to a pretrained model. 
% This modular perspective also reflects a broader shift in audio-language
% modeling toward explicit interfaces between continuous acoustic streams and
% language-model reasoning, as illustrated by recent work on always-on audio
% interaction~\citep{xie2026audiointeraction}.

Many prominent open music-generation projects release pretrained
weights and inference code but not the original training corpora or complete
training implementations
~\citep{yuan2026yue,yang2026heartmula,lei2026levo2,zhang2025inspiremusic,
yang2025songbloom}. Fully open alternatives demonstrate the value of
releasing data and training code, but may adopt architectures without a
separately trained diffusion renderer~\citep{jiang2026muse}. Consequently,
the effects of changing one module on the rest of an autoregressive
composition and diffusion-rendering pipeline remain difficult to measure.

Qwen-Music combines a 25~Hz single-codebook music tokenizer, an autoregressive
Music LLM, and a diffusion-based renderer
~\citep{xu2026qwenmusic}. Its authors report a 50.3\% preference rate against
Suno~V5.5 in a private bilingual blind test. 
The Qwen-Music report describes the high-level architecture and staged
training process. It does not release the training data, code, weights, or
many details needed for exact reproduction. Missing details include corpus
construction, token registration, optimization settings, sampling mixtures,
and checkpoint selection. Reconstructing the system therefore requires both
engineering choices and empirical validation.

We present Open-Qwen-Music, an open reconstruction of Qwen-Music and a
fully open research framework for this modular paradigm. It combines a
low-frame-rate semantic tokenizer, an
autoregressive planner, and a diffusion-based acoustic renderer under explicit
cross-module contracts: 25~Hz single-codebook Semantic Tokens, 6.25~Hz Melody
Tokens, and 48~kHz stereo output. The framework is trained with publicly
obtainable and locally generated data, including a synthetic corpus produced
with an open-weight MiniMax Music~3 model~\citep{minimax2026music3}. Beyond the reconstruction itself, the executable pipeline makes the interfaces among semantic representation, autoregressive generation, and acoustic rendering directly inspectable.

We ask two questions. First, how can a modular semantic-composition and
acoustic-rendering system be trained under explicit, independently
reproducible interfaces? Second, how can component-level diagnostics and artifact provenance make the behavior and limitations of a modular text-to-music system reproducible and inspectable?

Our main contributions are:

\begin{itemize}
    \setlength{\topsep}{0.35em}
    \setlength{\partopsep}{0pt}
    \setlength{\itemsep}{0.25em}
    \setlength{\parsep}{0pt}
    \setlength{\parskip}{0pt}
    \item We construct a complete modular system with explicit
    representation contracts, making the tokenizer, planner, spectral decoder,
    diffusion renderer, and refinement module independently trainable and
    replaceable.
    \item We report fixed-protocol evaluations of automatic supervision and
semantic discretization, qualitative end-to-end observations, and auditable
intermediate artifacts that support reproducible analysis of the released
system. The reported measurements characterize this implementation rather
than general properties of the architecture.
    \item We release the training datasets and provenance manifests,
    data-processing and annotation pipelines, complete training and inference
    implementations, evaluation code, configurations, and pretrained weights
    for every learned component at
    \url{https://github.com/biang15343100-source/Open-Qwen-Music}.
\end{itemize}

\section{Dataset and Data Pipeline}
\label{sec:dataset}

We organize the data into three nested corpora and one independent synthetic
set. Dataset-Inventory is the broad audio inventory. Dataset-Annotated is a deterministic,
source-capped subset of Dataset-Inventory. Dataset-Annotation-Filtered is an
annotation-filtered subset of Dataset-Annotated.
MiniMax is generated with MiniMax Music~3~\citep{minimax2026music3}.
Table~\ref{tab:training-corpora} gives their scale and use.

\begin{table}[H]
  \centering
  \footnotesize
  \caption{Dataset summary. Dataset-Annotation-Filtered is contained in Dataset-Annotated
  and is not counted as additional audio when both are used.}
  \label{tab:training-corpora}
  \begin{tabularx}{\textwidth}{@{}lrrY@{}}
    \toprule
    Dataset & Audio records & Audio hours & Definition and main use \\
    \midrule
    Dataset-Inventory
      & 1{,}481{,}301
      & 63{,}170.407
      & Broad 26-source inventory for tokenizer and Spec-VAE training. \\
    Dataset-Annotated
      & 240{,}325
      & 9{,}506.162
      & Source-capped, annotated subset of Dataset-Inventory. \\
    Dataset-Annotation-Filtered
      & 44{,}355
      & 2{,}155.176
      & Strict annotation-filtered subset of Dataset-Annotated: 35{,}484 vocal and
        8{,}871 instrumental records. \\
    MiniMax
      & 25{,}606
      & 1{,}103.057
      & Independent synthetic augmentation: 10{,}197 Chinese,
        9{,}955 English, and 5{,}454 instrumental records. \\
    Research package
      & 62{,}417
      & 2{,}958.571
      & Access-screened Dataset-Annotation-Filtered audio plus MiniMax audio;
        631 additional
        metadata-only records are not included in the audio count. \\
    \bottomrule
  \end{tabularx}
\end{table}

\subsection{Corpus Construction}
\label{sec:dataset-corpora}

\paragraph{Dataset-Inventory.}
Dataset-Inventory indexes 26 heterogeneous research, public-download, and synthetic
collections. Because these sources originate from independent open-data efforts, they carry no unified annotation schema---lyrics, section boundaries, and musical tags are absent, inconsistent, or source-specific. 
% It contains 1{,}431{,}451 training, 23{,}616 validation, and 26{,}234 test records. 
Records are variable-length source items; each model
applies its own eligibility rules and sampler. Appendix~\ref{app:data-sources}
lists every source, citation, and mirror or derivative caveat.

\paragraph{Dataset-Annotated.}
To enable the training of the Music LLM and the acoustic renderer, we developed a unified annotation pipeline that produces consistent lyrics, section structure, and musical tags across all sources. Annotating the entirety of Dataset-Inventory would be prohibitively expensive, so we cap each source at 20{,}000 records. The resulting corpus is Dataset-Annotated.

\paragraph{Dataset-Annotation-Filtered.}
Dataset-Annotation-Filtered is derived from Dataset-Annotated by applying strict annotation-quality filters and distributional balancing. On the quality side, vocal records require usable Melody-CoT labels, multiple sections, and reliable lyrics and tags; instrumental records must be non-synthetic and pass lyric, vocal-tag, and vocal-energy checks. Appendix~\ref{app:data-filtering} summarizes the main rules. On the balancing side, a deterministic weighted selection is followed by mixed-integer optimization that enforces a 4:1 vocal-to-instrumental ratio (consistent with Qwen-Music) and limits concentration by source, gender, language, and generator. The result contains 26{,}044 synthetic vocal, 9{,}440 non-synthetic vocal, and 8{,}871 non-synthetic instrumental records. It is drawn only from Dataset-Annotated's training split and has no separate validation or test split.

\paragraph{MiniMax generated set.}
We generate candidates with deterministic seeds, five target durations, and
at most five retries. Quality checks cover duration, loudness, clipping,
silence, DC offset, bandwidth, and spectral flatness. The retained audio is
44.1\,kHz, 16-bit PCM stereo. Generation conditions follow a controlled taxonomy.
\texttt{gemini-3.1-pro-preview} and \texttt{deepseek-v4-pro} provide lyrics
for the generation requests; these are not treated as transcripts. The
generated audio instead passes through the same ASR, alignment, and sectioning
pipeline as the other corpora; only style conditions are retained directly.

\paragraph{Training views.}
Dataset-Inventory supplies the broad candidate pool for the tokenizer and Spec-VAE.

The Music LLM training corpus contains 291{,}569 English
section-window records. Songs
are segmented using their annotated sections, with every training window
limited to 30 seconds, or 750 Semantic frames at 25 Hz. Its strict statistics
entry point recomputes 25-Hz training exposure and overlap-merged canonical
audio spans from that exact manifest.

\subsection{Preprocessing and Annotation}
\label{sec:data-pipeline}

Preprocessing validates source audio before deduplication and publication.
Annotation separates stems, infers structure, cross-checks two ASR
transcripts, aligns sections, estimates audio features and tags, and emits
eligibility flags. Only samples that pass validation enter downstream corpora.

\paragraph{Preprocessing.}
The preprocessing pipeline is index-only: it does not resample or copy source
audio. It records duration, sample rate, channel count, and eight quality
indicators. It rejects invalid durations, sample rates below 24\,kHz, bitrates
below 64\,kbps, fake stereo, severe bandwidth truncation, heavy boundary
silence, invalid channels, and extreme peaks. The accepted duration range is
5--360 seconds. Deduplication proceeds through URI, byte hash, external
identifier, and decoded-PCM fingerprint matching. These steps reduce
1{,}823{,}837 accepted parent records, totaling 83{,}022.76 hours, to
Dataset-Inventory.

\paragraph{Annotation.}
The annotation pipeline has 12 ordered stages. We separate vocals and
accompaniment with Hybrid Demucs~\citep{defossez2021hybrid}; the backend is
\texttt{HDEMUCS\_HIGH\_MUSDB\_PLUS}. SongFormer proposes song
sections~\citep{hao2025songformer}. Qwen3-ASR-1.7B transcribes both the vocal
stem and the mixture, and Qwen3-ForcedAligner-0.6B assigns timestamps
~\citep{shi2026qwen3asr}. We cross-check both transcripts before section
fusion. Recent work on recognition under compound real-world distortions
shows that weakened acoustic grounding can produce omissions and
hallucinations~\citep{xie2026megaasr}; accordingly, both transcription paths
are treated as uncertain supervision rather than human ground truth.

We then estimate voice activity, pitch, and spectral properties. pYIN is used
for annotation-time pitch analysis~\citep{mauch2014pyin}; the separate Melody
Tokenizer uses RMVPE, as described in Section~\ref{sec:music-llm}. A controlled
Qwen3-Omni-30B-A3B-Instruct tagger produces genre, mood, instrument,
vocal-gender, and vocal-timbre labels plus a short description
~\citep{xu2025qwen3omni}. The final stage combines this evidence into
task-specific eligibility flags.

% \paragraph{Splits and leakage controls.}
% Duplicate groups in Dataset-Inventory are assigned to train, validation, and test with
% target probabilities of 96\%, 2\%, and 2\%; no group crosses a split. A
% separate 2{,}372-record tokenizer test set is checked against V2 training data
% by UID, content identity, and variant group. For legacy V1, we apply the same
% checks where linking is possible and exclude unresolved records. These
% controls prevent known exact or grouped overlap, but do not prove removal of
% all acoustic near-duplicates or external-benchmark contamination.

\paragraph{Annotation limits.}
All annotations---lyrics, sections, tags, descriptions---are produced by off-the-shelf open-source models used without any training or fine-tuning. They are therefore automatic estimates, not human ground truth, and may contain systematic or per-sample errors. Confidence thresholds serve only as filters, not accuracy estimates. Dataset-Annotation-Filtered is not an acoustic-fidelity set, so Render uses a separate acoustic gate. 
%We omit corpus-wide language, style, and vocal proportions for
%Dataset-Inventory and Dataset-Annotated because no reliable aggregate audit is available.
%BPM, key, and mode

\subsection{Research Data Release}
\label{sec:data-release}
To let the community build on our work without repeating the costly
annotation pipeline, we publicly release the two highest-quality corpora
from our pipeline---Dataset-Annotation-Filtered and the MiniMax generated set---after applying
a permission-aware desensitization process.
The pinned research package is available at
\url{https://huggingface.co/datasets/david-miller-45678/open-qwen-music-dataset-c-minimax-music3}.
It combines 36{,}811 access-screened Dataset-Annotation-Filtered audio records with all 25{,}606
MiniMax records. We exclude 6{,}912 records from
the Jamendo-QA mirror~\citep{archit2026jamendoqamirror,koh2025jamendoqa},
MoisesDB~\citep{pereira2023moisesdb},
MuChin-v2~\citep{karlwang2025muchinv2,wang2024muchin}, and
Music4All~\citep{santana2020music4all} because those sources require
permission for access or redistribution.
Cambridge-MT~\citep{seniorCambridgeMT} contributes 631 metadata-only records.
One JamendoMaxCaps record is quarantined by the privacy scanner.
The artifact manifest records source-level access and redistribution
decisions separately for audio, annotations, tokens, latents, and model
weights, and the public package contains only artifacts covered by the
corresponding release policy.

% The package contains 827 uncompressed WebDataset shards plus JSONL and Parquet
% metadata. Each sample carries a public ID, audio hash, provenance, rights,
% annotations, and eligibility flags. Manifests and SHA-256 checksums bind the
% release. The privacy gate removes internal paths, contact details,
% credential-like strings, and private container fields; 172 containers are
% remuxed without re-encoding.

% Because 5{,}038 records carry explicit Creative Commons NonCommercial terms,
% the package is limited to non-commercial academic research under
% \texttt{LicenseRef-OQM-Academic-\allowbreak Research-Only-1.0}; commercial
% training and product use are prohibited. Source-specific rights still apply,
% and unclear records remain marked. Users must inspect per-record rights before
% reuse or redistribution. The package is not uniformly open-licensed.

\section{Model and Training}
\label{sec:model-training}

The system reconstructs Qwen-Music's tokenizer, autoregressive Music LLM, and
diffusion Render modules~\citep{xu2026qwenmusic}. Table~\ref{tab:contracts}
fixes their shared interfaces; changing any entry invalidates downstream
assets and checkpoints.

\begin{table}[H]
  \centering
  \small
  \caption{Cross-module interfaces.}
  \label{tab:contracts}
  \begin{tabular}{@{}llll@{}}
    \toprule
    Representation & Rate & Specification & Downstream role \\
    \midrule
    Semantic Tokens & 25\,Hz & one codebook, 32{,}768 & Music LLM and DiT \\
    Melody Tokens & 6.25\,Hz & 256; unvoiced $=255$ & Melody-CoT \\
    Spectral latent & 25\,Hz & 128 dimensions & Spec Decoder \\
    Waveform & 48\,kHz & stereo & playback and evaluation \\
    \bottomrule
  \end{tabular}
\end{table}

\subsection{Music Tokenizer}
\label{sec:music-tokenizer}

\paragraph{Architecture.}
The tokenizer maps 24\,kHz mono audio to a 25\,Hz stream from one
32{,}768-entry codebook, a nominal rate of 375\,bit/s. A 128-bin log-Mel
frontend ($n_{\mathrm{fft}}=2048$, hop 240, $f_{\max}=12$\,kHz) runs at
100\,Hz. A causal ConvNeXt frontend (kernel 7, expansion 4) reduces the rate
by four before a 24-layer, 1024-wide Conformer with 16 attention heads, a
4096-dim feed-forward network, and RoPE. A single 16-dimensional cosine
vector quantizer is inserted after layer~13; three lightweight heads predict
multilingual CTC subwords, log-Mel, and soft Chroma. Deployment retains only
the causal encoder up to the quantizer. The frontend and all Conformer
convolutions are causal in every stage; self-attention becomes strictly
causal during the causal-adaptation and subsequent phases.

Qwen-Music specifies the broad architecture and four training stages but
omits most feature, optimization, and budget settings. The choices below are
our reconstruction decisions, not claims about the official implementation.

\paragraph{Tokenizer training recipe.}
Table~\ref{tab:tokenizer-recipe} summarizes the five functional phases used
to train the semantic tokenizer.

\begin{table}[H]
  \centering
  \small
  \caption{Training recipe for the semantic tokenizer.}
  \label{tab:tokenizer-recipe}
  \begin{tabularx}{\textwidth}{@{}llY@{}}
    \toprule
    Phase & Objective & Main reconstruction choice \\
    \midrule
    Pretraining & Bidirectional BestRQ
       & Broad pretraining with waveform masking and four-frame targets. \\
    Domain adaptation & BestRQ
       & Fresh optimizer and sampler-matched feature statistics. \\
    Causal adaptation & Causal BestRQ
       & Strictly causal attention; unchanged targets. \\
    Semantic supervision & Multi-task prediction
       & 4096-unit multilingual CTC, Mel, and Chroma losses with weights
         $1/1/1$. \\
    Discretization & Vector quantization
       & 16-dimensional cosine VQ, data-dependent initialization, exact
         all-rank usage statistics, and no dead-code revival. \\
    \bottomrule
  \end{tabularx}
\end{table}

The first phase learns broad music representations without lyric labels through
BestRQ masked prediction: 0.4\,s waveform spans (40 frames) are masked with
probability 0.3 and perturbed by $-20$\,dB relative-RMS noise, and the
four-frame target is quantized against a frozen $8192\times16$
random-projection codebook. Training uses deterministic 30-second crops from
Dataset-Inventory under a source-balanced sampler. Domain adaptation then uses the annotated
domain on 175{,}204 records (8{,}509 hours) from Dataset-Annotated and MiniMax with
a fresh optimizer. Causal adaptation makes self-attention strictly causal so the
encoder matches its autoregressive downstream use, using a global
active-frame micro-average; targets are unchanged.

Semantic supervision first trains a new
4{,}096-unit multilingual subword CTC head alone, then thaws the backbone
with CTC, Mel, and Chroma losses at equal weights (signed L1 for
Mel and Chroma), using complete tracks up to 300 seconds and subwords rather
than characters to respect the 25\,Hz CTC capacity.
This use of recognition supervision at a discrete bottleneck is closely
related to ASR-guided speech content quantization, where the bottleneck is
trained to preserve linguistic information~\citep{liu2025asrqvc}.
The final phase inserts the
cosine quantizer (code dimension 16, commitment $\beta=0.25$, EMA codebook
decay 0.99) and trains with CTC/Mel/Chroma/VQ weights
$0.5/1/1/1$. The codebook is initialized from 1{,}048{,}576 real frames via
PCA whitening and spherical $k$-means; a gate warmup and a frozen-lower phase
ease the continuous-to-discrete transition, using exact all-rank usage
statistics with no dead-code revival.

\subsection{Music LLM and Melody-CoT}
\label{sec:music-llm}

\paragraph{Reconstruction scope.}
Qwen-Music autoregressively predicts Semantic Tokens from musical tags and
structured lyrics, optionally after generating a Melody-CoT plan
~\citep{xu2026qwenmusic}. Its reported 3B Qwen3.5-Omni initialization is not
publicly available. We use the public Qwen2.5-Omni-3B thinker
~\citep{qwen25omni2025}, preserving model scale but not architecture. A
deterministic serializer replaces the unavailable Prompt Rewriter. Owing to
limited data and compute, our LLM is pretrained only on a fixed English
corpus of section-based windows no longer than 30 seconds, without curriculum
learning.
No post-training is performed---no
supervised fine-tuning, offline DPO, or online GSPO---unlike Qwen-Music,
which reports a full multi-stage post-training pipeline.

\paragraph{Sequences and token registry.}
Each record contains structured music tags, ordered song sections, lyrics,
Semantic Tokens, optional section-aligned Melody Tokens, and provenance. We
implement the three published layouts:
\[
\begin{aligned}
\mathbf{s}_{\mathrm{plain}} &=[\mathbf{x},\mathbf{z}],\\
\mathbf{s}_{\mathrm{section}} &=
    [\mathbf{x},\mathbf{m}_{\mathrm{section}},\mathbf{z}],\\
\mathbf{s}_{\mathrm{unique}} &=
    [\mathbf{x},\mathbf{m}_{\mathrm{unique}},\mathbf{z}],
\end{aligned}
\]
where $\mathbf{x}$ is the serialized text condition, $\mathbf{m}$ is Melody-CoT,
and $\mathbf{z}$ is the Semantic Token sequence. Section mode includes one
melody segment for every lyric-bearing section. Unique-section mode groups
segments by section label and samples one representative per label anew in
each epoch.  The implementation retains all three layouts, including Melody-CoT, while the Music LLM finally uses the direct text-to-semantic layout (i.e., plain mode). The two Melody-CoT layouts remain available as alternative interfaces.

Token-level coupling between an internal plan and an acoustic response has
also been explored in large speech models~\citep{xie2025miniomnireasoner}; our
layouts instantiate the related planning--generation interface for music,
while final inference uses the direct layout described above.

Melody Tokens follow the Qwen-Music rule. RMVPE first estimates pitch
~\citep{rmvpe2023}. After reduction to 50\,Hz, eight-frame median pooling
produces a 6.25\,Hz sequence. Voiced frames encode a clipped MIDI offset from
the median voiced pitch; token 255 denotes unvoiced frames.

A token registry assigns disjoint global ranges to text, control,
Semantic, and Melody tokens; the vocabulary is padded to a fixed size. Render
consumes local Semantic IDs in $[0,32768)$, never the expanded global IDs.

\paragraph{Objective and context.}
 Final training
uses a 4{,}096-token per-GPU budget and at most 750 Semantic frames (30
seconds) per section-based example. The released inference configuration uses
the same explicit Semantic-frame budget. We optimize next-token
cross-entropy on Melody (optionally) and Semantic positions while masking the text
condition, with no auxiliary loss or label smoothing.

\paragraph{Data and optimization.}
 The training corpus contains 291{,}569 English
windows. The windowing
procedure follows annotated section boundaries and limits every example to
30 seconds. Its three quality strata are sampled with fixed weights 0.31479,
0.12719, and 0.55802.

\begin{table}[H]
  \centering
  \small
  \caption{Optimization recipe for the autoregressive planner.}
  \label{tab:music-llm-recipe}
  \begin{tabular}{@{}lc@{}}
    \toprule
    Setting & Music LLM \\
    \midrule
    Initialization & Qwen2.5-Omni thinker \\
    Learning rate / schedule & $1\times10^{-4}$ / constant \\
    Parallelism & 64 H20 GPUs, DDP, BF16 \\
    Context & 4{,}096 tokens; $\leq$30 s / 750 Semantic frames \\
    \bottomrule
  \end{tabular}
\end{table}

%3{,}532{,}894{,}208-parameter
The planner is trained using DDP on 64 H20 GPUs, BF16 computation, AdamW at a constant $10^{-4}$ learning rate, gradient clipping at 2, and two-step gradient accumulation.

\paragraph{Decoding configuration.}
The decoder uses temperature 0.7,
top-$p$ 0.98, Semantic repetition penalty 0.1, and a 64-token repetition
window.

\subsection{Acoustic Render}
\label{sec:render}

\paragraph{Published design and scope.}
Qwen-Music Render turns frame-aligned Semantic Tokens and text conditions into
audio: a 1.3B conditional Diffusion Transformer maps the conditions to a
128-dimensional, 25\,Hz acoustic latent, a spectral VAE decodes the latent
into a coarse 48\,kHz stereo complex spectrogram, and a Band-Mode Refiner
corrects frequency-band residuals before inverse STFT~\citep{xu2026qwenmusic}.
The paper discloses the backbone geometry (32 blocks, 1024 hidden width, 24
attention heads, expansion 8, self-attention $\to$ cross-attention $\to$
feed-forward order, timestep AdaLN, RoPE), the text encoders (a frozen
Qwen3-Embedding-0.6B model for descriptions, a six-layer transformer for
lyrics, text-only dropout for CFG), the VAE interfaces ($[2,480,T]$ complex
spectra over a SpectroStream-style seven-block encoder/decoder with
delayed-fusion stereo), and the Refiner's outer training recipe. It omits the
STFT configuration, posterior form, DiT head geometry, semantic fusion
operator, flow path and timestep distribution, solver, CFG, and EMA settings,
and most optimization budgets. The values below are our reconstruction
decisions, not claims about the official implementation.

\paragraph{Training recipe.}
Table~\ref{tab:render-recipe} lists the Render components and their training
objectives. As with the tokenizer, every model artifact is bound by SHA-256 in
the artifact manifest.

\begin{table}[H]
  \centering
  \small
  \caption{Training recipe for the acoustic renderer.}
  \label{tab:render-recipe}
  \begin{tabularx}{\textwidth}{@{}llY@{}}
    \toprule
    Component & Objective & Main reconstruction choice \\
    \midrule
    Spec-VAE & Reconstruction
       & Full VAE on 48\,kHz stereo; global batch 256. \\
    Spec-VAE & Decoder fine-tuning
       & Encoder frozen; STFT-domain adversarial training at batch 64. \\
    DiT & Conditional flow matching
       & 24-block, width-1536 backbone; exponential-moving-average
         weights. \\
    Refiner & Band-aware residual correction
       & Frozen VAE with fresh optimizers; mid-band magnitude and phase correction. \\
    \bottomrule
  \end{tabularx}
\end{table}

\paragraph{Spec-VAE.}
Our Spec-VAE preserves the published interfaces. Audio enters at 48\,kHz
stereo through an STFT with $n_{\mathrm{fft}}=960$ and an equal window length,
hop 480, no centering, and a 240-sample left pad; DC is retained and Nyquist
dropped, giving 480 frequency bins per channel and the $[2,480,T]$
representation. The encoder fuses stereo channels late and the decoder splits
them early, both in the published seven-block 2D convolutional style. The
posterior is a diagonal Gaussian: training draws posterior samples, while
validation reconstruction uses the mean. Initial training fits the full VAE
on Dataset-Inventory at batch 256. Decoder fine-tuning then freezes the encoder and
uses batch 64 while adding an STFT-domain
adversarial objective (multi-resolution STFT, a mixed-scale spectral term,
and waveform adversarial and feature-matching terms).

\paragraph{Conditional DiT.}
The renderer uses a Stable Audio 3-compatible profile: 24
blocks, hidden width 1536, 24 attention heads of
dimension 64, feed-forward multiplier 4, QK RMS normalization with
differential attention, and 64 learned memory tokens. Local, global, and text
conditioning each use width 1024; Semantic Tokens remain aligned at 25 Hz to
the 128-dimensional spectral latent.

The model is trained by conditional flow matching with the VAE encoder frozen.

Training uses 64 H20 GPUs, batch one per rank,
four-step gradient accumulation (global parent batch 256), BF16, AdamW at
$5\times10^{-5}$ with betas 0.9/0.95 and weight decay 0.01, a brief warmup
followed by a constant rate, gradient clipping at 1.0, and FP32 exponential
moving averages with decay 0.999. A global loudness condition uses
$-15.4335$ LUFS .

\paragraph{EMDC.}
Because the Music LLM generates rather than copies Semantic Tokens, the
DiT's training conditions are cleaner than those it sees in deployment. We
adapt the EMDC strategy of FullDiT~\citep{fullsong}: after clean training,
each semantic frame is independently replaced with probability
$p_{\mathrm{replace}}$ by a code sampled from the $K$ cosine-nearest
neighbors of its ID in the tokenizer's 16-dimensional codebook. The
replacement probability, neighborhood size, and schedule are configuration
parameters recorded in the configuration files; optimizer, moving-average
policy, data, and VAE remain unchanged during this adaptation.

\paragraph{Band-Mode Refiner.}
The Refiner sits between the spectral decoder and inverse STFT, using a
12-layer ConvNeXt-1D trunk (width 256, kernel 7) with hard band boundaries
at 1.5 and 4\,kHz: low band phase only, mid band magnitude and phase, high
band magnitude only. Output projections are zero-initialized so the untrained
network is a learnable identity (our interpretation of the published zero
initialization), and phase errors use wrapped-L1 instantaneous-frequency and
group-delay terms. Refiner training freezes the VAE and uses fresh optimizers
under the published outer recipe.

\paragraph{Inference configuration.}
The inference pipeline chains the diffusion renderer, spectral VAE,
and band-aware Refiner, with an 8-step Euler
solver, CFG 1.0, and global loudness $-15.4335$ LUFS. The artifact manifest
binds these artifacts and their configurations.

\section{Evaluation}
\label{sec:evaluation}

We report held-out quantitative evaluations of automatic supervision and
semantic discretization, followed by a qualitative listening assessment of
the complete generation pipeline. Each quantitative table states its
evaluation set, sample count, aggregation, and metric scale.

\subsection{Annotation Quality}
\label{sec:annotation-quality}

The training pipeline relies on automatically generated lyrics and musical
attributes. 
%We audit this supervision on 299 held-out Chinese songs. Of these, 298 produce evaluable annotations and one fails during upstream decoding.
Table~\ref{tab:annotation-quality} reports the measurements for which
the audit protocol is fixed.

\begin{table}[H]
  
  \centering
  \small
  \caption{Automatic-annotation audit. Each row retains its own task-specific
  denominator and is not combined into an overall annotation score.}
  \label{tab:annotation-quality}
  \begin{tabularx}{\textwidth}{@{}lYl@{}}
    \toprule
    Signal & Result & Note \\
    \midrule
    Lyric LCS recall & 0.889 & precision 0.879; CI $[0.877, 0.902]$ \\
    Vocal gender & 88.3\% & CI $[85.9, 90.3]$ \\
    Genre & 75.2\% & GTZAN subset \\
    Mood & 76.9\% & arousal after dead-zone mapping \\
    Confidence--error correlation & Spearman $-0.764$ &
      two-transcription confidence \\
    \bottomrule
  \end{tabularx}
\end{table}

The negative confidence--error correlation indicates that agreement between
the two transcription paths is informative for data-quality filtering.
Because the rows measure different annotation tasks and use different
reference subsets, we interpret them individually rather than averaging them.

\subsection{Semantic Tokenizer Quantization}
\label{sec:tokenizer-quantization-diagnostics}

We measure the information loss introduced by assigning continuous features
to discrete codes. The same examples and fixed prediction heads are applied
to the 16-dimensional projected representation ($P$) and its nearest-code
quantized counterpart ($Q$). Table~\ref{tab:tokenizer-quantization} reports
the increase in prediction error from $P$ to $Q$; lower values indicate
better preservation.

\begin{table}[H]
  
  \centering
  \small
  \caption{Quantized-minus-continuous ($Q-P$) prediction-error differences.
  PER differences are percentage points. Mel was not measured on the common
  music holdout.}
  \label{tab:tokenizer-quantization}
  \begin{tabular}{@{}lrrrr@{}}
    \toprule
    Evaluation panel & Examples & $\Delta$PER (pp) & $\Delta$Mel & $\Delta$Chroma \\
    \midrule
    Common music holdout & 2{,}372 & 2.352 & ---    & 0.00124 \\
    Exposure panel       & 256   & 2.153 & 0.0306 & 0.00159 \\
    Real-recording panel & 101   & 2.879 & 0.0267 & 0.00143 \\
    \bottomrule
  \end{tabular}
\end{table}

Approximately 98.2\% of the 32{,}768-entry codebook is active across the
evaluated panels, indicating no evident codebook collapse on these data.
Qwen-Music reports more than 99\% utilization under its own protocol
~\citep{xu2026qwenmusic}; because utilization depends on the corpus and the
number of observed frames, the values provide context rather than a
controlled ranking.

The 2.15--2.88 percentage-point PER increase quantifies the additional
phonetic prediction error associated with nearest-code assignment. Mel and
Chroma differences provide complementary within-system diagnostics. No
external result uses the same paired $P\rightarrow Q$ protocol, so these
measurements are not used for cross-system ranking.

\subsection{Qualitative Listening Observations}
\label{sec:end-to-end-subjective-evaluation}

We qualitatively assess two paths through the frozen generation pipeline.
The demos are shown at \url{https://huggingface.co/spaces/oqmtest1451/open-qwen-music-demo}.
This assessment records recurring listening observations rather than
numerical preference estimates.

\paragraph{Reconstruction from reference-audio Semantic Tokens.}
Semantic Tokens extracted from a source recording are passed directly to the
acoustic renderer. The resulting audio generally preserves the sung lyrics:
most words remain intelligible and follow the source content. The main audible
limitations are an electronic coloration in the vocal timbre and a stereo
image with limited width and depth.

\paragraph{End-to-end text-to-music generation.}
The Music LLM first generates Semantic Tokens from the text and lyric
conditions, after which the same acoustic renderer synthesizes the waveform.
In this setting, some lyric words or phrases are omitted, while others are
rendered indistinctly. Relative to reconstruction from reference-audio Semantic Tokens, this
pattern suggests that errors in the Music-LLM-generated semantic sequence
are an important contributor to the additional degradation observed in
the rendered output.

\section{Discussion}
\label{sec:conclusion}

\paragraph{Summary.}
Open-Qwen-Music reconstructs the Qwen-Music semantic-to-waveform pipeline
under a constrained but fully specified setting: approximately 10{,}100 hours
of training exposure compared with the more than five million hours reported
by Qwen-Music, one 64-GPU budget per module, automatically generated
annotations, and public replacements for unavailable components
~\citep{xu2026qwenmusic,minimax2026music3,qwen25omni2025}. Within these
constraints, the system completes the full path from text and lyrics to
48\,kHz stereo audio. The purpose of the reconstruction is to make the
architecture reproducible and inspectable rather than to claim quality parity
with Qwen-Music. We therefore release the training resources, model weights,
and data-processing, training, inference, and evaluation pipelines needed to
reproduce the system.

\paragraph{Perceptual characterization.}
Qualitative listening separates two sources of degradation. When the renderer
is conditioned on reference-audio Semantic Tokens, most lyric content remains
intelligible, showing that the acoustic stack can express vocal content from a
reliable semantic sequence. The reconstructed vocals nevertheless retain an
electronic coloration, and the stereo image has limited width and depth. In
end-to-end text-to-music generation, where the Music LLM supplies the Semantic
Tokens, generation quality is less stable and vocals are frequently missing
or indistinct. This contrast suggests that the quality of Music-LLM-generated Semantic
Tokens is an important contributor to failures in vocal intelligibility.
The renderer also limits timbral and spatial fidelity, and the present
qualitative comparison does not isolate the relative causal contributions
of the two stages.

\paragraph{Main problems.}
The current system has several limitations.
\begin{itemize}
  \item \emph{Automatic supervision remains imperfect.} The annotation audit
  reports lyric LCS recall of 0.889 and attribute accuracies between 75.2\%
  and 88.3\%. Errors in lyrics and musical attributes can propagate into the
  tokenizer, Music LLM, and renderer conditions.
    \item \emph{Semantic discretization introduces measurable information
loss.} The codebook is broadly active, with 98.2\% utilization on the
evaluated panels, while nearest-code assignment raises PER by
2.15--2.88 percentage points relative to the projected continuous
representation. The projected features are normalized before cosine
quantization. This design may improve codebook coverage while discarding
magnitude-related information, and it may increase the difficulty of
downstream sequence modeling. Establishing this causal link requires
additional evidence, such as conditional-entropy estimates, Music-LLM
negative log-likelihood, or a controlled comparison between tokenizer
variants.
  \item \emph{Music-LLM-generated Semantic Tokens are an important
contributor to vocal degradation.}
  The model is trained on English section windows no longer than 30 seconds.
  In qualitative end-to-end generation, its Semantic Tokens lead to unstable
  output and frequent missing or indistinct vocals, while an independent
  held-out Music LLM evaluation is not yet available.
  \item \emph{Acoustic rendering limits fidelity.} With reference-audio Semantic
  Tokens, the renderer generally preserves lyrics but retains electronic vocal
  coloration and weak stereo width and depth.
  \item \emph{Evaluation coverage remains limited.} Current quantitative
  results cover annotation quality and tokenizer quantization. The system-level
  listening observations are qualitative, and no matched comparison with
  another text-to-music system is reported.
\end{itemize}

\paragraph{Planned improvements.}
These limitations define the next iteration of the project.
\begin{itemize}
  \item \emph{Data.} Expand the training corpora while preserving item-level
  provenance and redistribution controls, improve the automatic annotation
  models, and introduce human-verified subsets for annotation calibration.
  \item \emph{Representation.} Reduce the information lost at the semantic
  bottleneck and evaluate tokenizer changes by their effect on downstream
  generation as well as by component-level diagnostics.
  \item \emph{Music LLM.} Improve the stability and completeness of generated
  Semantic Tokens, then evaluate the model on an independent held-out set with
  lyric coverage, termination, and semantic consistency measures. Evaluation
  will also extend to additional languages and longer cross-section contexts.
  \item \emph{Post-training.} Curate high-quality supervision and investigate
multi-objective alignment across musicality, lyric intelligibility, and
control adherence, using rewards computed from rendered audio rather than
token sequences alone. This direction is motivated by both multi-objective
LLM alignment and perceptual-feedback optimization in generative speech,
where token-level objectives need not track perceptual quality
~\citep{li2025multiobjective,li2025perceptualspeech}.
  \item \emph{Rendering.} Reduce electronic vocal coloration, improve stereo
  width and depth, and strengthen robustness to Music-LLM-generated Semantic
  Tokens while retaining the lyric preservation observed with
  reference-audio Semantic Tokens.
  \item \emph{Evaluation.} Add multi-seed generation, controlled human
  listening, and protocol-compatible comparisons with public text-to-music
  systems.
\end{itemize}

\paragraph{Open Release and Outlook.}
The remaining gap is not a missing pipeline component; it lies in data scale
and supervision quality, in the Semantic Token representation formed by the
tokenizer, in the acoustic and spatial fidelity of the renderer, and in the
current evaluation coverage. In particular, normalization and quantization discard part of the input
information and may increase the difficulty of modeling the resulting
Semantic Token sequences; this hypothesis requires controlled tokenizer
comparisons or direct sequence-modeling measurements.
Although the present results do not establish quality parity with Qwen-Music,
we release the system in full. To our knowledge, Open-Qwen-Music is the first
fully open release of a text-to-music system that combines LLM-based semantic
composition with diffusion-based acoustic rendering and makes every layer of
the research stack public: training datasets with provenance manifests,
data-processing and annotation pipelines, model definitions, training and
inference code, evaluation code and configurations, and pretrained weights
for every learned module. We intend it as a reproducible foundation for research on this architecture
and broader audio-language modeling~\citep{xie2026audiointeraction}, and will
continue to improve its generation quality, controllability, and robustness.

\bibliographystyle{tmlr}
\bibliography{reference}

\clearpage
\appendix

\section{Detailed Data Information}
\label{app:data}

\subsection{Dataset-Inventory Source Breakdown}
\label{app:data-sources}

\begin{table}[H]
  \centering
  \scriptsize
  \caption{The 26 retained sources in Dataset-Inventory. Hours are rounded to two
  decimals. Each row cites the primary paper, the ingested release page, or
  both. Derivative and mirror caveats are stated below.}
  \label{tab:dataset-inventory-sources}
  \begin{tabularx}{\textwidth}{@{}Yrr@{}}
    \toprule
    Source & Records & Hours \\
    \midrule
    AI-Music Deduplicated~\citep{aimusic2026deduplicated}
      & 508{,}557 & 25{,}141.23 \\
    JamendoMaxCaps~\citep{roy2025jamendomaxcaps}
      & 261{,}739 & 13{,}683.13 \\
    Muse~\citep{jiang2026muse}
      & 117{,}067 & 7{,}589.09 \\
    Udio Dataset~\citep{blanchon2024udio}
      & 134{,}329 & 3{,}914.82 \\
    FMA Full~\citep{defferrard2017fma}
      & 65{,}282 & 3{,}607.52 \\
    GlobalDISCO~\citep{solak2025globaldisco}
      & 73{,}024 & 3{,}340.90 \\
    MTG-Jamendo Raw 30s~\citep{bogdanov2019mtg}
      & 45{,}342 & 2{,}636.36 \\
    ACE-Step Songs~\citep{yi2025acestepsongs,jiang2025mufun}
      & 21{,}568 & 1{,}012.61 \\
    Music4All~\citep{santana2020music4all}
      & 95{,}699 & 797.49 \\
    Cambridge-MT~\citep{seniorCambridgeMT}
      & 6{,}856 & 409.86 \\
    Jamendo-QA mirror~\citep{archit2026jamendoqamirror,koh2025jamendoqa}
      & 6{,}396 & 394.31 \\
    MuChin-v2~\citep{karlwang2025muchinv2,wang2024muchin}
      & 4{,}216 & 269.85 \\
    AnyInstruct Music~\citep{zhan2024anygpt}
      & 118{,}368 & 166.37 \\
    MoisesDB~\citep{pereira2023moisesdb}
      & 1{,}348 & 79.00 \\
    OpenMIC-2018~\citep{humphrey2018openmic}
      & 16{,}789 & 46.64 \\
    MAESTRO v3.0.0~\citep{hawthorne2019maestro}
      & 367 & 24.10 \\
    MERGE Bimodal~\citep{louro2024mergedataset,louro2026merge}
      & 1{,}570 & 13.07 \\
    MusicNet~\citep{thickstun2017musicnet}
      & 130 & 8.92 \\
    MUSDB18-HQ~\citep{rafii2019musdb18hq}
      & 140 & 8.59 \\
    PMEmo2019~\citep{zhang2019pmemorelease,zhang2018pmemo}
      & 780 & 8.24 \\
    MedleyDB-derived mix~\citep{amaailabMedleyDB,bittner2014medleydb,
      bittner2016medleydb2}
      & 129 & 6.78 \\
    Coimbra MIREX~\citep{panda2013multimodal}
      & 726 & 5.98 \\
    GuitarSet~\citep{xi2018guitarset}
      & 360 & 3.05 \\
    URMP~\citep{li2019urmp}
      & 42 & 1.28 \\
    MusicCaps raw-audio mirror~\citep{google2023musiccaps,
      agostinelli2023musiclm}
      & 253 & 0.70 \\
    Song Generation Binisha~\citep{binisha2024songgeneration}
      & 224 & 0.51 \\
    \bottomrule
  \end{tabularx}
\end{table}

The source labels identify the artifacts ingested by our pipeline, not
necessarily the official upstream distributions. In particular, Jamendo-QA is
the \texttt{Archit00} mirror. The PMEmo row uses the updated 2019 release. The
MedleyDB row is an AMAAI Lab derivative containing instrument/other mixes,
rather than the official multitracks. MusicCaps raw audio comes from a local
third-party mirror containing 326 candidates before filtering, whereas the
official MusicCaps release contains 5{,}521 metadata records. Its citations
identify the upstream benchmark, not an independently archived provenance
record for that audio mirror.

\subsection{Dataset-Annotation-Filtered Eligibility Rules}
\label{app:data-filtering}

\begin{table}[H]
  \centering
  \small
  \caption{Main Dataset-Annotation-Filtered eligibility rules. The implementation also applies
  silence, timeline, and annotation-consistency checks.}
  \label{tab:dataset-annotation-filtered-rules}
  \begin{tabularx}{\textwidth}{@{}lYY@{}}
    \toprule
    Branch & Required evidence & Main numeric rules \\
    \midrule
    Vocal
      & Melody-CoT eligible; at least two sections; no under-transcribed,
        dropped, or hallucinated lyric windows; English, Chinese, or
        bilingual.
      & Lyric confidence $\geq0.905$; lyric coverage $\geq0.15$; genre
        $\geq0.80$; instrument $\geq0.75$; binary vocal gender
        $\geq0.89$. \\
    Instrumental
      & Non-synthetic; no usable lyrics or vocal tags; fewer than three text
        units.
      & Instrumental-gender confidence $\geq0.90$; vocal ratio $<0.15$;
        vocal-stem-to-mix ratio $<-15$\,dB. \\
    \bottomrule
  \end{tabularx}
\end{table}

Lyric confidence thresholds for broader task eligibility are 0.723 for
pretraining, 0.905 for Melody-CoT, and 0.945 for SFT. Tags are retained at
confidence 0.40 or above. These are fixed filtering thresholds rather than
estimates of human-label accuracy; the SFT threshold is reserved for future
supervised fine-tuning data curation.

\subsection{Annotation Pipeline}
\label{app:annotation-pipeline}

The released annotation pipeline is a dependency-aware directed acyclic graph
with 12 stages. Independent evidence streams are preserved until fusion: the
two transcription passes are produced separately, and generative tag
predictions are retained separately from acoustic voice evidence. Table
\ref{tab:annotation-pipeline} summarizes the input-output role of each stage.

\begin{table}[H]
  \centering
  \scriptsize
  \caption{Stages of the released annotation pipeline. The order follows the
  dependency graph; stages whose prerequisites are satisfied may run in
  parallel.}
  \label{tab:annotation-pipeline}
  \begin{tabularx}{\textwidth}{@{}rlY@{}}
    \toprule
    & Stage & Function \\
    \midrule
    1 & Index
      & Ingest source manifests or file trees, assign stable sample identities,
        and record source and audio metadata. \\
    2 & Source separation
      & Extract the vocal stem and vocal-activity evidence used by the lyric and
        voice-analysis branches. \\
    3 & Structure analysis
      & Detect musical-section boundaries and normalize predicted section labels
        to the shared section taxonomy. \\
    4 & Vocal-stem transcription
      & Produce the primary lyric transcription from the separated vocal stem. \\
    5 & Mixture transcription
      & Independently transcribe the original mixture for cross-checking the
        vocal-stem result. \\
    6 & Lyric fusion
      & Reconcile the two transcription streams with available source text,
        determine language, and produce the selected lyric sequence with quality
        measurements. \\
    7 & Lyric alignment
      & Assign word- or character-level timestamps to the selected lyric
        sequence. \\
    8 & Section assembly
      & Combine structural boundaries, aligned lyrics, and vocal activity into
        section-level records and structured lyrics. \\
    9 & Voice acoustics
      & Extract pitch and spectral evidence from the vocal stem for the vocal
        attributes used during tag fusion. \\
    10 & Generative tagging
      & Produce a free-form description and candidate genre, mood, instrument,
        vocal-gender, and vocal-timbre tags under controlled output taxonomies. \\
    11 & Tag fusion
      & Normalize tag candidates and fuse generative, acoustic, and structural
        evidence while retaining per-attribute confidence and source information. \\
    12 & Validation and export
      & Assemble the final annotation record, apply confidence gates and
        downstream-use eligibility rules, validate the schema, and emit the
        training manifest. \\
    \bottomrule
  \end{tabularx}
\end{table}

\subsection{Annotation Output Schema}
\label{app:annotation-schema}

Each exported row follows a single versioned record schema. Downstream corpus
construction reads this record rather than stage-specific intermediate files.
Table~\ref{tab:annotation-schema} lists the public field groups and their
semantics.

\begin{table}[H]
  \centering
  \scriptsize
  \caption{Field groups in each exported annotation record.}
  \label{tab:annotation-schema}
  \begin{tabularx}{\textwidth}{@{}lY Y@{}}
    \toprule
    Field group & Contents & Role \\
    \midrule
    Identity
      & Schema version, stable sample identifier, and source-dataset label.
      & Joins annotations to the corresponding corpus item while preserving the
        record contract. \\
    Audio
      & Audio reference, duration in seconds, sample rate, and channel count.
      & Defines the media item and its technical properties. \\
    Language
      & Resolved lyric language, when available.
      & Supports language-aware normalization, filtering, and corpus sampling. \\
    Description
      & Free-form musical description produced by the tagging branch.
      & Provides text conditioning beyond the controlled tag vocabulary. \\
    Musical tags
      & Genre, mood, instrument, vocal gender, and vocal timbre; low-confidence
        attributes are omitted.
      & Supplies independently controllable prompt attributes without treating
        missing evidence as a negative label. \\
    Structured lyrics
      & Lyrics serialized in section order with section labels.
      & Provides the text and musical-form representation consumed by downstream
        sequence construction. \\
    Sections
      & Section identifier and label, start and end time, lyric and vocal flags,
        section text, aligned-unit count, and lyric coverage.
      & Preserves the time-aligned relationship among form, vocals, and lyrics. \\
    Provenance and quality
      & Lyric-source status and confidence, alignment availability, tag
        confidence and evidence sources, and processing-completeness indicators.
      & Makes filtering decisions traceable without requiring intermediate-stage
        outputs. \\
    Usability
      & Eligibility flags and reasons for supported downstream training uses.
      & Allows each training corpus to apply its own evidence requirements from
        the same annotation record. \\
    \bottomrule
  \end{tabularx}
\end{table}

\section{Training and Inference Configuration}
\label{app:training-configuration}

This section consolidates the optimization and decoding settings used for the
reported system. The tables separate the sequential training phases and report
effective global batch sizes after gradient accumulation.

\subsection{Semantic Tokenizer}

All six optimization phases use 64 GPU processes, one gradient-accumulation
step, BF16 computation, and gradient clipping at 1.0. AdamW uses
$\beta_1=0.9$ and $\beta_2=0.98$; the learning rate is linearly warmed up and
then cosine-decayed to one tenth of its peak value. Weight decay is 0.01 except
during supervised-head warmup, where it is zero.

\begin{table}[H]
  \centering
  \scriptsize
  \caption{Sequential optimization phases for the semantic tokenizer. Batch/GPU
  is the per-process batch size; warmup and updates count optimizer steps.}
  \label{tab:appendix-tokenizer-optimization}
  \begin{tabularx}{\textwidth}{@{}lYrrrr@{}}
    \toprule
    Phase & Objective and training unit & Updates & Batch/GPU & Global batch & Peak LR / warmup \\
    \midrule
    Masked pretraining
      & Bidirectional BestRQ on deterministic 30-s crops
      & 100{,}000 & 2 & 128 & $2\!\times\!10^{-4}$ / 10{,}000 \\
    Music-domain adaptation
      & BestRQ on deterministic 30-s crops from the annotated domain
      & 20{,}000 & 4 & 256 & $2\!\times\!10^{-5}$ / 500 \\
    Causal adaptation
      & Causal BestRQ with unchanged acoustic targets
      & 20{,}000 & 4 & 256 & $2\!\times\!10^{-5}$ / 500 \\
    Supervised-head warmup
      & Frozen backbone; CTC head on complete tracks up to 300 s
      & 2{,}000 & 1 & 64 & $1\!\times\!10^{-3}$ / 100 \\
    Joint semantic supervision
      & CTC, Mel, and Chroma losses with weights $1/1/1$
      & 40{,}000 & 1 & 64 & $5\!\times\!10^{-5}$ / 3{,}000 \\
    Discrete codebook training
      & CTC/Mel/Chroma/VQ weights $0.5/1/1/1$
      & 25{,}000 & 1 & 64 & $3\!\times\!10^{-5}$ / 2{,}000 \\
    \bottomrule
  \end{tabularx}
\end{table}

Masked pretraining uses 0.4-s waveform spans with masking probability 0.3 and
$-20$-dB relative-RMS noise. The final three phases use a 4{,}096-unit
multilingual CTC vocabulary; Mel and Chroma use signed L1 losses. Before the
last phase, the $32{,}768\times16$ cosine codebook is initialized from
1{,}048{,}576 projected frames by PCA whitening and five iterations of
mini-batch spherical $k$-means. The quantizer uses commitment weight 0.25, a
5{,}000-step gate warmup, and a 5{,}000-step frozen-lower-encoder interval.

\subsection{Music LLM}

The Music LLM is optimized only in the direct text-to-semantic sequence layout.
Both stages use 64 GPU processes, one sequence per process, two gradient-
accumulation steps, BF16 computation, and an effective global batch size of
128. Stage~2 continues the data cursor from Stage~1 but starts a new optimizer.

\begin{table}[H]
  \centering
  \small
  \caption{Two-stage Music LLM optimization. Cumulative updates count the
  uninterrupted sequence of consumed training batches.}
  \label{tab:appendix-llm-optimization}
  \begin{tabularx}{\textwidth}{@{}lYrrr@{}}
    \toprule
    Stage & Initialization and data progression & Stage updates & Cumulative updates & Global batch \\
    \midrule
    Stage 1 & Qwen2.5-Omni-3B initialization; beginning of the corpus stream
      & 5{,}000 & 5{,}000 & 128 \\
    Stage 2 & Stage-1 model initialization; continuation of the corpus stream
      & 5{,}000 & 10{,}000 & 128 \\
    \bottomrule
  \end{tabularx}
\end{table}

Both stages use AdamW with learning rate $1\times10^{-4}$, betas
$0.9/0.95$, weight decay 0.01, a constant schedule without warmup, and
gradient clipping at 2.0. The per-GPU token budget is 4{,}096, with at most
750 Semantic frames per example. Training uses 291{,}569 English windows of at
most 30 s, sampled from the three data strata with probabilities 0.31479,
0.12719, and 0.55802.

\subsection{Acoustic Models}

All acoustic modules operate on 48-kHz stereo audio. The Spec-VAE and Refiner
use 1.28-s random crops; the Renderer predicts 128-dimensional VAE latents at
25 Hz. Table~\ref{tab:appendix-acoustic-schedules} gives the effective
distributed schedules.

\begin{table}[H]
  \centering
  \scriptsize
  \caption{Distributed training schedules for the acoustic modules.}
  \label{tab:appendix-acoustic-schedules}
  \begin{tabularx}{\textwidth}{@{}lYrrrrrl@{}}
    \toprule
    Module or stage & Objective & Updates & GPUs & Batch/GPU & Accum. & Global batch & Precision \\
    \midrule
    VAE Stage 1 & Full-model reconstruction and KL training
      & 90{,}000 & 64 & 4 & 1 & 256 & FP32 \\
    VAE Stage 2 & Decoder-only adversarial refinement
      & 5{,}000 & 64 & 1 & 1 & 64 & FP32 \\
    Renderer & Conditional flow matching
      & 29{,}000 & 64 & 1 & 4 & 256 & BF16 \\
    Refiner & Band-aware residual correction
      & 1{,}250 & 64 & 1 & 1 & 64 & FP32 \\
    \bottomrule
  \end{tabularx}
\end{table}

\paragraph{Spec-VAE.}
The common analysis STFT uses a 960-sample Hann window, a 480-sample hop, no
centering, and a 240-sample left pad; DC is retained and Nyquist is removed.
Stage~1 uses Muon at $1\times10^{-4}$ with momentum 0.95, weight decay 0.1,
and a constant learning rate. Its objective combines multi-resolution STFT
loss (weight 1), KL regularization ($10^{-6}$), and mixed-scale spectral loss
($2\times10^{-6}$). Stage~2 freezes the encoder and uses separate Muon
optimizers for the decoder and discriminator at $3.75\times10^{-5}$ and
$3.75\times10^{-6}$, respectively. Both use momentum 0.95; the learning rate
warms up for 1{,}000 steps and cosine-decays to 10\% of its peak. Stage~2 adds
waveform adversarial and feature-matching losses with weights 0.03 and 3.0.

\paragraph{Renderer.}
The Renderer uses linear rectified flow, a Gaussian source, uniformly sampled
timesteps, velocity prediction, and uniform loss weighting. AdamW uses learning
rate $5\times10^{-5}$, betas $0.9/0.95$, weight decay 0.01, a 2{,}000-step
warmup followed by a constant rate, and gradient clipping at 1.0. FP32
exponential-moving-average parameters use decay 0.999. The first 25{,}000
updates use clean Semantic Tokens, followed by a 1{,}000-update corruption
ramp; the remaining updates use the full mixture, in which 25\% of examples
receive codebook-neighbor corruption with $K=384$.

\paragraph{Refiner.}
The Refiner keeps the VAE fixed and alternates four generator updates per
discriminator update. AdamW learning rates are $1.5\times10^{-4}$ for the
generator, $1\times10^{-6}$ for the waveform discriminator, and
$1\times10^{-5}$ for the spectral discriminator; all use betas $0.9/0.95$
and weight decay 0.01. The objective weights are 0.5 for STFT reconstruction,
0.1 for instantaneous-frequency/group-delay reconstruction, 0.175 and 20 for
waveform adversarial and feature matching, and 0.5 and 35 for spectral
adversarial and feature matching.

\subsection{Inference}

\begin{table}[H]
  \centering
  \small
  \caption{Inference settings used for text-to-music generation.}
  \label{tab:appendix-inference-settings}
  \begin{tabularx}{\textwidth}{@{}lY@{}}
    \toprule
    Setting & Value \\
    \midrule
    Music LLM sequence layout & Direct text-to-semantic \\
    Sampling & temperature 0.7; top-$p$ 0.98; top-$k$ 0 \\
    Semantic repetition control & penalty 0.1 over a 64-token window \\
    Acoustic solver & Euler, 8 steps \\
    Classifier-free guidance & 1.0 \\
    Loudness condition & $-15.4335$ LUFS \\
    Output & 48-kHz stereo waveform \\
    \bottomrule
  \end{tabularx}
\end{table}

\end{document}